\documentclass[reprint,amsmath,amssymb,aps,prb,superscriptaddress]{revtex4-1}
\usepackage{amsmath}
\usepackage{amssymb}
\usepackage{dcolumn}
\usepackage{graphicx}
\usepackage{longtable}
\usepackage{hyperref}
\usepackage{bm}
\usepackage{color}
\usepackage{epsfig}
\usepackage{epstopdf}
\usepackage[utf8]{inputenc}
\usepackage[T1]{fontenc}
\usepackage{mathptmx}
\usepackage{adjustbox}
\usepackage{multirow}
\usepackage{comment}

\newcommand{\seebeckunit}{$\mu$VK$^{-1}$}
\newcommand{\PFunit}{$\mu$Wcm$^{-1}$K$^{-2}$}
\newcommand{\sigmaunit}{$\Omega^{-1}$m$^{-1}$}
\newcommand{\kappaunit}{Wm$^{-1}$K$^{-1}$}

\newcommand{\zt}{$ZT$}

\begin{document}
	
	\title{Rhodium based half-Heusler alloys as possible  optoelectronic and thermoelectric materials}
	
	\author{Dhurba R. Jaishi}
	
	\author{Sujit Bati}
	
	\author{Nileema Sharma}
	
	\author{Bishnu Karki}
	
	\author{Bishnu P. Belbase}

	\author{Madhav Prasad Ghimire}
	\email{madhav.ghimire@cdp.tu.edu.np}
	\affiliation{Central Department of Physics, Tribhuvan University, Kirtipur 44613, Kathmandu, Nepal \&}
	\affiliation{Condensed Matter Physics Research Center (CMPRC), Butwal 32907, Rupandehi, Nepal}
		
\date{\today}
\begin{abstract}
 On the basis of density functional theory and semi-classical Boltzmann theory, we have investigated the structural, elastic, electronic, optical and thermoelectric properties of 18--valence electron count rhodium based half-Heusler alloys focusing on RhTiP, RhTiAs, RhTiSb, and RhTiBi. The absence of imaginary frequencies in the phonon dispersion curve for these system verifies that they are structurally stable. RhTiP is ductile in nature, while others are brittle. The alloys are found to be semiconducting with indirect band gaps ranging from 0.94 to 1.01 eV. Our calculations suggest these materials to have high absorption coefficient and optical conductivity in the ultraviolet as well as visible region. While considering thermoelectricity, we found that $p$--type doping is more favorable in improving the thermoelectric properties. The calculated values of power factor with $p$-type doping are comparable to some of the reported half-Heusler materials. The optimum figure of merit \zt\ is $\sim1$ for RhTiBi suggesting it as a promising candidate for thermoelectric applications while RhTiP, RhTiAs, and RhTiSb with optimum \zt \ values between 0.38 to 0.67 are possible candidates for use in thermoelectric devices.
\end{abstract}
	
\maketitle

\section{Introduction}

In recent years, research on materials science are found to focus in search for novel materials suitable for applications in optoelectronics, spintronics, thermoelectrics, etc. For these, various compounds including oxides, perovskites, Heusler alloys, chalcogenides, and many other groups were explored to tune their properties.\cite{hu2012half,leskela2019atomic, ghimire2010study, gh42015half, kumar2020topological,ghimire2017chemical,roknuzzaman2017towards}
The growing demand for energy and over reliance on fossil fuels have alerted the scientific community to seek alternative sources of energy. Another crucial aspect to look into for energy conservation is to reduce power loss. A significant portion of total energy is wasted in form of unusable heat every year.\cite{ovik2016review}  Thus, technology to recuperate some of the wasted energy can be very helpful in solving the problem of energy shortage and escalating energy costs. This can also help in reducing the emission of greenhouse gases caused by the excessive use of fossil fuels. One of the techniques used to generate electricity from waste-heat generated from vehicles, factories, etc. is the thermoelectrics (TE). \cite{rowe2006thermoelectrics,snyder2008,sootsman2009new, tritt2000recent}

The suitability of materials for TE application is measured in terms of a dimensionless figure of merit, $ZT = \sigma \alpha^2T/\kappa$, where $\sigma$ is the electrical conductivity, $\alpha$ is the Seebeck coefficient, $T$ is the temperature, and $\kappa$ is the thermal conductivity. The thermal conductivity consists of the contribution from the lattice thermal conductivity ($\kappa_l$)  and the electronic thermal conductivity ($\kappa_e$). Enhancement of TE efficiency of the system can be done by tailoring these interrelated parameters. The value of \zt \ can be improved through reduction of the lattice thermal conductivity and increasing the power factor, and this can be achieved through various means like doping, band engineering, and reducing dimension of the materials.\cite{poudel2008high, mao2017phonon, zhao2016enhanced, pei2012band,neophytou2011effects}

Among others, half-Heusler (hH) compounds are found to be mechanically stable and possess magnificent thermal stability and large power factor.\cite{he2016achieving, rogl2016mechanical,rausch2015long} These features make hH alloys suitable for high temperature TE applications. Remarkably, many doped hH alloys are found to have a good TE properties with \zt \ values that are comparable to  the bismuth, lead, silicon-germanium containing conventional TE  materials such as Bi$_2$Te$_3$, BiSbTe, PbTe, BiSb, and SiGe.\cite{goldsmid1954use,poudel2008high,gelbstein2005high,poudeu2006nanostructures,lee2012large} However, high values of lattice thermal conductivity limits the \zt \ values for undoped hH alloys. Recently, $p$--type FeNbSb,\cite{fu2015realizing}, ZrCoBi,\cite{zhu2018discovery} and TaFeSb\cite{zhu2019discovery}--based hH compounds have been explored as TE materials with high \zt \ $\sim$ 1.5. This high value of \zt \ is due to the reduction of thermal conductivity with the help of dopants which cause more phonon scattering, while maintaining high carrier mobility. Many other experimentally and theoretically identified hH alloys show interesting TE  properties.\cite{kim2007high,yu2009high,chen2013effect,fu2014high,
rai2015study,PhysRevMaterials.1.074401,zeeshan2017Co,zeeshan2018fetasb,singh2019first,
ning2020high,dhurba2021,hori2020} 
Although doping has been the most widely used method to improve \zt \ and power factor of hH alloys, application of pressure has also been shown to be fruitful in improving the TE properties.\cite{kumarasinghe2019band, ning2020high} At present, it is well known that hH alloys are excellent candidates for use in TE devices, but some of them require expensive dopants to simultaneously tune the carrier concentration and suppress the lattice thermal conductivity. Therefore, further research is required to identify efficient, non-toxic, low-cost, and naturally abundant multifunctional materials.

Apart from TE, the hH alloys have drawn significant attention due to the possibility of tunable band gap,\cite{kumarasinghe2019band} high optical absorption, optical conductivity in the broad range of incident spectrum etc.\cite{wei2018thermoelectric,kieven2010ii}  The noticeable band gap becomes useful in optoelectronic applications, and they further act as a potential source in replacement of Pb containing alloys.\cite{eperon2015inorganic,ramasamy2016all, akkerman2016strongly} 

From our literature survey, we noticed very few theoretical works on rhodium based hH alloys.\cite{anissa2019optical, bamgbose2021exploring, azin2021comparison}  The reported work is preliminary and needs extensive investigation to understand the electronic, optical, mechanical, and thermoelectric properties of these materials. With this motivation, we explore RhTiZ (where Z are P, As, Sb, and Bi) which consists of two transition elements - Rh and Ti. The localized $d$ orbitals of these elements are expected to contribute heavy bands near the Fermi level ($E_{\textrm F}$)\cite{yang2008evaluation} which can be beneficial for TE properties as their Seebeck coefficient usually increases with increase in the effective mass of the charge carriers.

\section{COMPUTATIONAL DETAILS}
\begin{figure}[hb]
	\centering
	\psfig{figure=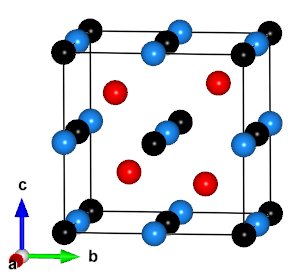,scale=0.5}
	\caption{The crystal structure of cubic hH RhTiZ (Z = P, As, Sb, Bi). The balls in red, blue, and black color represent Rh, Ti, and $Z$ atoms respectively. The structure was drawn  using VESTA.\cite{momma2011vesta}}
	\label{Fig1}
\end{figure}

The hH compounds RhTiZ (Z = P, As, Sb, Bi) belong to the cubic MgAgAs$-$type structure with space group $F\bar{4}3m$. The crystal structure of RhTiZ is shown in Figure \ref{Fig1}. In this structure Ti and $Z$ form rock salt structure, with $Z$ and either of the Rh or Ti forming zinc-blende structure. The Rh, Ti, and $Z$ atoms occupy the Wyckoff position 4c (1/4, 1/4, 1/4), 4b (1/2,1/2,1/2), and 4a (0,0,0) respectively.
The structure optimization, electronic structure, mechanical stability, and  optical property calculations are performed using the linearized augmented plane-wave (LAPW) method based on density functional (DF) theory as implemented in the WIEN2k code.\cite{blaha2001wien2k} We used the basis set cut-off parameter $RK_{max}$ = 7, and the muffin-tin radii (RMTs) were taken in the ranges 2.0--2.5 Bohr for all the atoms. The standard generalized gradient approximation (GGA)\cite{perdew1996generalized} along with the modified
Becke$-$Johnson (mBJ) potential\cite{PhysRevLett.102.226401} were used. A well converged $k-$mesh of $12 \times12 \times12$ was used for the self-consistent calculations and structure optimization. A denser $k-$mesh of  $50\times50 \times50$ was used for the optical and transport properties calculations. To check the phonon stability, we performed the phonon band structure calculation using Quantum ESPRESSO (QE) package\cite{giannozzi2009quantum} based on DF perturbation theory on a 4$\times$4$\times$4 \textbf{\textit{q}}--mesh with the norm-conserving Perdew-Zunger exchange-correlation functional. The electronic transport properties are calculated by solving the semi-classic Boltzmann transport equation (BTE). The calculation is performed by using the BoltzTraP2 code.\cite{BoltzTraP2} Though preliminary work was reported earlier,\cite{azin2021comparison} $\kappa_l$ has not been considered properly involving lattice vibrations and phonon dispersion.  In order to determine the accuracy of \zt, values of $\kappa$ is essential. Thus, the lattice thermal conductivity, $\kappa_l$, is evaluated here by solving the linearized BTE within the single-mode relaxation time approximation (SMA) using thermal2 code.\cite{fugallo2013ab,fugallo2014thermal,cepellotti2015phonon}

\section{RESULTS AND DISCUSSION}
\subsection{Structure optimization, mechanical and phonon stability}
\begin{figure}[h!]
	\centering
	\psfig{figure=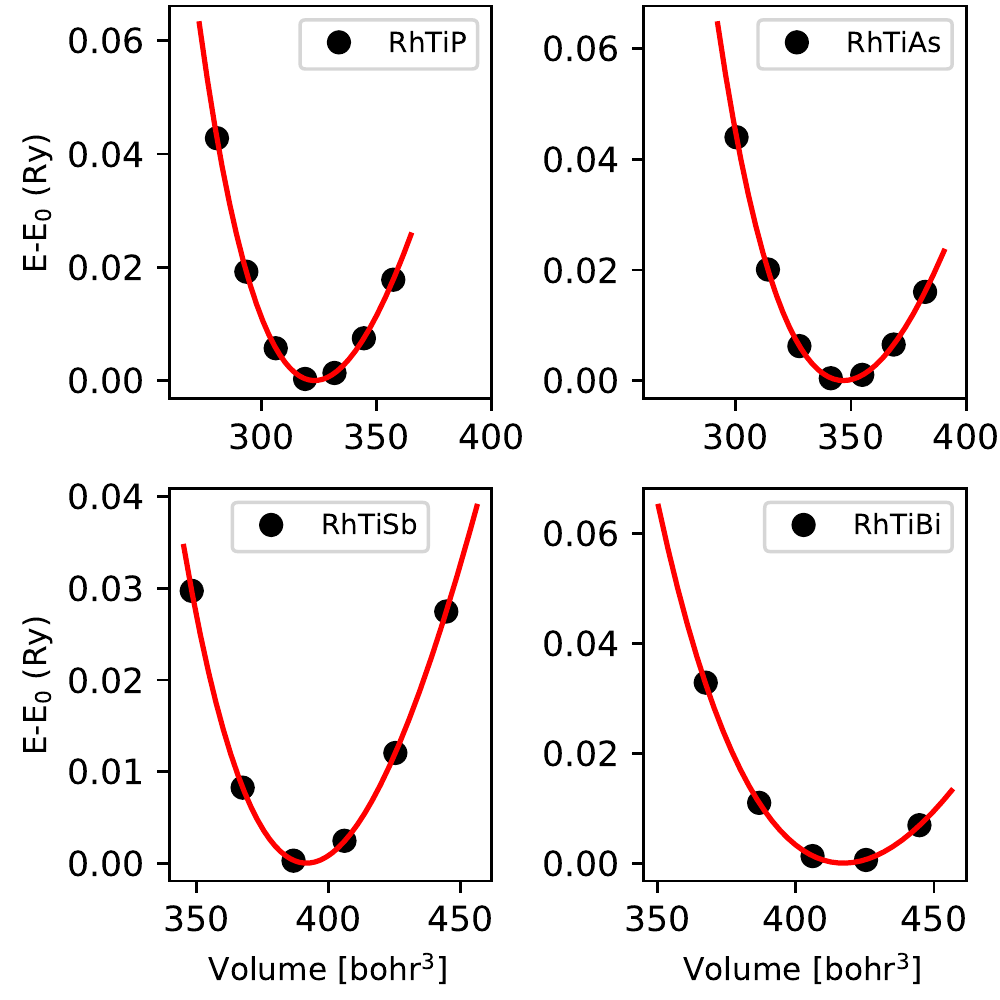,width=\columnwidth}
	\caption{E--V diagram for RhTiP, RhTiAs, RhTiSb, and RhTiBi. Here E$_0$ is the ground state energy corresponding to optimized volume.}
	\label{Fig}
\end{figure}

The lattice constants obtained by optimizing the volume and corresponding band gaps using GGA and mBJ functionals are listed in Table \ref{tab1}. The energy \textit{vs.} volume (E--V) plot for the RhTiP, RhTiAs, RhTiSb, and RhTiBi are shown in Figure \ref{Fig}. The optimized volumes of the unit cells are obtained by using Murnaghan equation of state.\cite{murnaghan1944compressibility} The optimized lattice constants are found consistent with the previous theoretical results.\cite{ma2017computational}

\begin{table}[ht]
	\caption{The optimized lattice constant $a$ and the band gap $E_{g}$ within GGA and mBJ for the cubic hH alloys RhTiZ (Z = P, As, Sb, Bi).}
	\begin{tabular}{ccccccccccccc}
		\hline\hline
		&&&&&&&GGA&&mBJ\\
		\hline
		System & & & & a (\AA) & & & E$_g$(eV) & &E$_g$(eV)& \\
		\hline\\
		RhTiP  & & & & 5.76 & & & 0.84& & 1.01 \\ \\
		RhTiAs & & & & 5.90 & & & 0.85& & 1.06 \\ \\
		RhTiSb & & & & 6.14 & & & 0.77& & 0.94 \\ \\
		RhTiBi & & & & 6.27 & & & 0.73& & 0.94 \\
		\hline\hline
	\end{tabular}
	\label{tab1}
\end{table}

Mechanical stability of the crystal is checked by calculating the elastic constants on the basis of the well-known Born stability criteria\cite{born1940stability}: C$_{11}$-C$_{12}$ $>$ 0, C$_{11}$+2C$_{12}$ $>$ 0, C$_{44}$ $>$0. The independent elastic constants C$_{11}$, C$_{12}$, and C$_{44}$ for cubic hH compounds calculated by using the finite strain theory\cite{murnaghan1937finite} and are presented in the Table \ref{tab2}. The calculated values are positive and thus shows that RhTiZ is mechanically stable. Using the second order elastic constants, bulk modulus ($B$), shear modulus ($G$), Young’s modulus ($Y$), Pugh's ratio (i.e., the ratio of the bulk and shear modulus) ($B/G$), and Poisson ratio ($\nu$) of the hH alloys are calculated as tabulated in Table \ref{tab2}.
\begin{table}[ht]
	\caption{Calculated elastic constants C$_{ij}$, bulk modulus (B), shear modulus (G), Pugh's ratio (B/G), Young's modulus (Y), Poisson ratio ($\nu$), hardness (H), volume (V), density ($\rho$), transverse sound velocity ($\nu_t$), longitudinal sound velocity ($\nu_l$), average sound velocity ($\nu_m$), and Debye temperature ($\Theta_D$) for the cubic hH alloys RhTiZ (Z = P, As, Sb, Bi). The unit of $B$, $G$, $Y$, and H is GPa.}
	\begin{tabular}{ccccccccccccc}
		\hline
		& &RhTiP &RhTiAs &RhTiSb& RhTiBi \\
	\hline\hline\\
		$C_{11}$&&201.41&305.85&243.93&250.72& \\ \\
		$C_{12}$&&160.87&86.71 &96.67 &66.08& \\ \\
		$C_{44}$&&107.62&176.03&137.78&123.68& \\ \\
		\hline
		$B$&&174.38&159.76&145.75&127.63& \\ \\
		$G$&&56.09&145.55&107.15&110.01& \\ \\
		$B/G$&&3.12&1.09&1.36&1.16& \\ \\
		$Y$ &&151.99&334.94&258.18&156.37& \\ \\
		$\nu$&&0.35&0.15&0.20&0.17& \\ \\
		$H$ &&5.629&33.979&21.515&14.701\\\\
		\hline
		$V$ (\AA$^3$)&&191.55&205.51&232.15&247.15 \\ \\
		$\rho$(g/cm$^3$) && 6.30&7.29 &7.79 & 9.67 \\ \\
		$\nu_t$ (m/s)&&2983.51 &4467.21&3707.23&3373.41 \\ \\
		$\nu_l$ (m/s)&&6288.25&6965.06&6084.34&5326.89\\ \\
		$\nu_m$  (m/s)&&3356.55&5906.76&4094.82&3710.74 \\ \\
		$\Theta_D$ (K)&&397.07&567.00&454.34&403.22 \\ 
		\hline
	\end{tabular}
	\label{tab2}
\end{table}
\begin{figure}[h!]
	\centering
	\psfig{figure=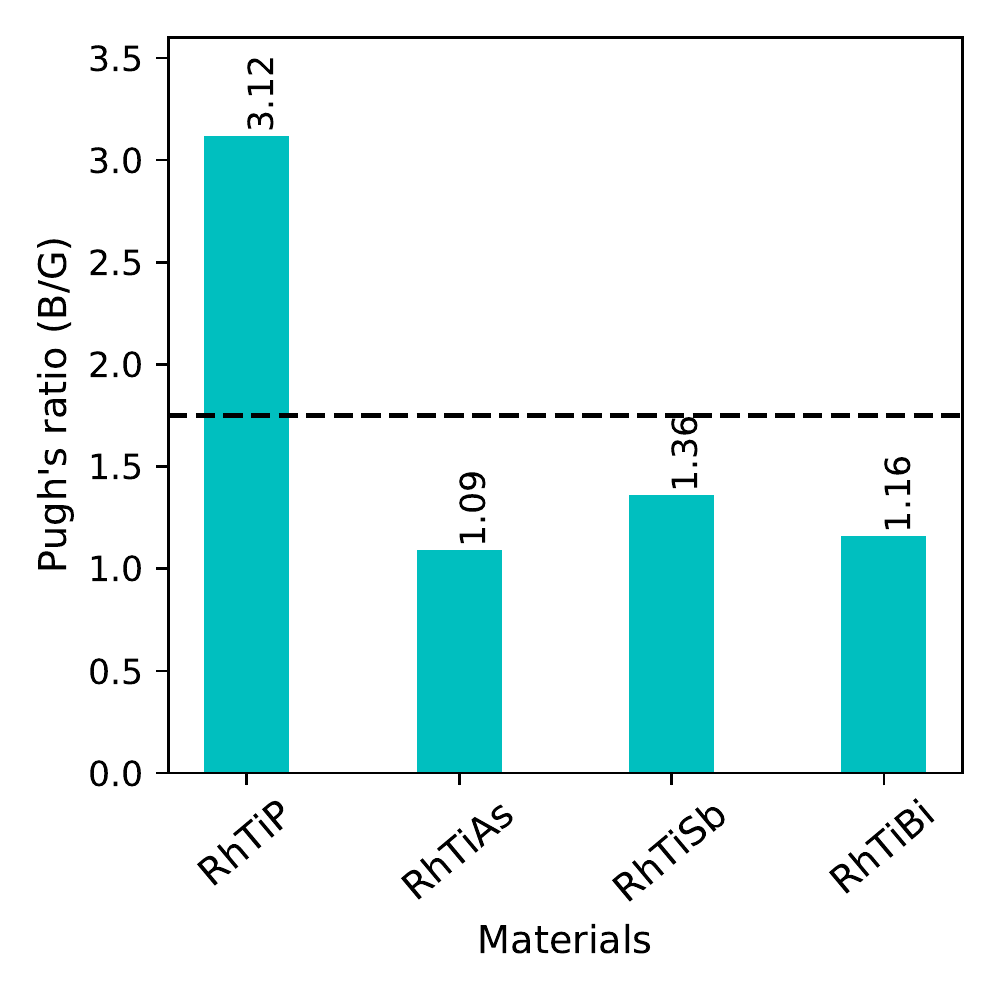,width=3in, height=3in}
	\caption{Variation of the Pugh’s ratio of the cubic RhTiZ (Z = P, As, Sb, Bi) hH materials. The horizontal dashed line separates the ductile materials from brittle.}
	\label{Fig2}
\end{figure}
The bulk modulus is a measure of degree of stiffness of the material. The calculated values of bulk modulus of hH alloys lies between 127.63 GPa (for RhTiBi) to 174.38 GPa (for RhTiP). Upon substitution of different atoms (say lighter P to heavier Bi) at the $Z-$sites, the value of bulk modulus is found to decrease. This indicates an increase in flexibility of RhTiZ as $Z$ is replaced by P, As, Sb, and Bi respectively. The shear and Young's modulii are maximum for RhTiAs. 
The Pugh's ratio and Poisson ratio determines the ductility and brittleness of the material. The critical value of Pugh's ratio and Poisson ratio are 1.75 and 0.26, \cite{frantsevich1982elastic} respectively, that separates the ductile material from the brittle one. If the value of Pugh's ratio is greater than 1.75 or when the Poisson ratio is greater than 0.26, the material is ductile or else the material is brittle in nature. Under this criteria, RhTiP is a ductile material while the other three hH alloys are brittle in nature. This property not only measures the strength in response to mechanical shocks but also ease in manufacturing. The enhancement of mechanical strength makes it more durable. The plot of Pugh's ratio for these four compounds are shown in Figure \ref{Fig2}. 
We further consider the hardness ($H$) which can be derived by using Poisson's ratio and Young's modulus. As seen from Table \ref{tab2}, RhTiAs has the highest hardness while RhTiP has the least. Here, we also calculate the longitudinal and transverse acoustic wave velocities, average sound velocity and Debye temperature\cite{schreiber1973elastic,anderson1963simplified} shown in Table \ref{tab2}. This information is important for the synthesis and design of these materials. The $\Theta_D$ obtained at 0 K are 397.07, 567.00, 454.34, and 403.22 for RhTiP, RhTiAs, RhTiSb, and RhTiBi, respectively. The calculated melting temperature \cite{fine1984elastic} of the system ranges from 1740 -- 2360 ($\pm$ 300) K. The obtained values of $\nu_m$, $\Theta_D$, and high value of melting temperature shows that these hH alloys will have lower lattice thermal conductivity and high temperature stability. These features indicates that RhTiZ can be a suitable for TE applications.
\begin{figure}[h!]
	\centering
	\psfig{figure=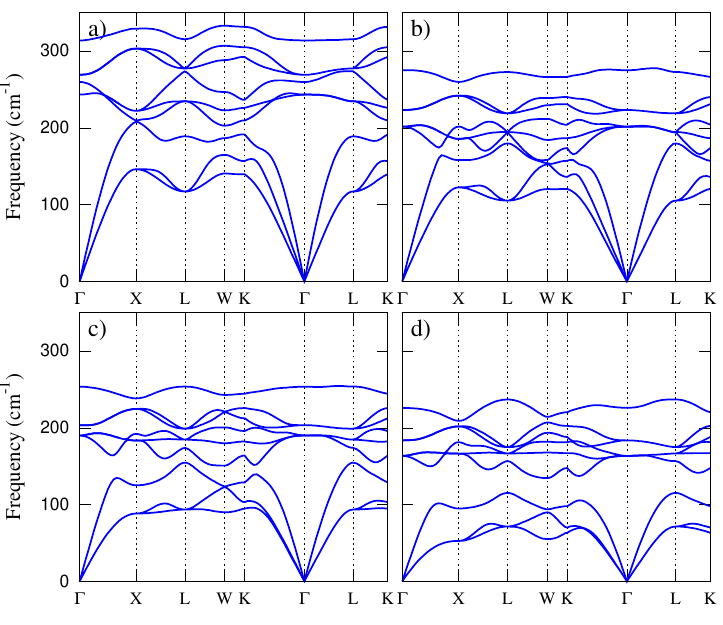,width=\columnwidth}
	\caption{Phonon dispersion curves for (a) RhTiP, (b) RhTiAs, (c) RhTiSb, and (d) RhTiBi.}
	\label{Fig3}
\end{figure}

The calculated phonon dispersion of hH materials are shown in Figure \ref{Fig3}. This measures the dynamical stability of the materials. The absence of phonon modes with imaginary frequencies suggests that these alloys are dynamically stable. As hH alloys are composed of three types of atoms, it comprises a total of nine phonon modes: three acoustic modes  at lower-frequency region and six optical modes at the high-frequency region. The maximum acoustic phonon mode of RhTiP lies around 200 cm$^{-1}$ which decreases as the element P is replaced by As, Sb, or Bi. In  case of RhTiBi the acoustic modes lie within frequency range $\sim$ 100 cm$^{-1}$. This indicates that phonon group velocity are in the order: RhTiP $>$ RhTiAs $>$ RhTiSb $>$ RhTiBi.

\subsection{Electronic structure}
\begin{figure}[h!]
	\centering
	\psfig{figure=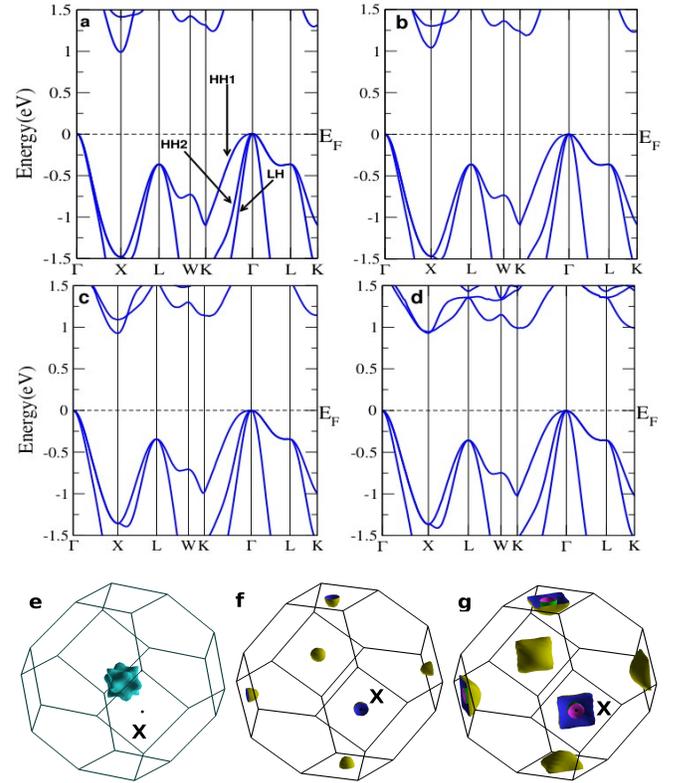,width=\columnwidth}
	\caption{Calculated electronic band structure for  (a) RhTiP, (b) RhTiAs, (c) RhTiSb, and (d) RhTiBi within mBJ and iso-energy surfaces (e) at 4 meV below VBM for RhTiBi, (f) at 4 meV above CBM for RhTiAs, and (g) at 4 meV above CBM for RhTiBi. In (f) and (g), blue and violet colors are the inner surface, whereas yellow and green colors are the outer surface.}
	\label{band}
\end{figure}

The calculated band structure of RhTiZ (Z = P, As, Sb, Bi) along the high symmetry directions in the first Brillouin zone are shown in Figure \ref{band}. The materials are found to be semiconducting with indirect band gaps ranging from $0.94-1.01$ eV within mBJ calculations. The values are comparable with the earlier works. \cite{anissa2019optical,bamgbose2021exploring, azin2021comparison} 
The band gap variation is small between the valence band maximum (VBM) and conduction band minimum (CBM). To note is, the band next to CBM moves close to it when P is replaced by As, Sb, or Bi (see Figure \ref{band}). The reason for this is due to the extended nature of the band from the heavier elements.

\begin{figure}[h!]
	\centering
	\psfig{figure=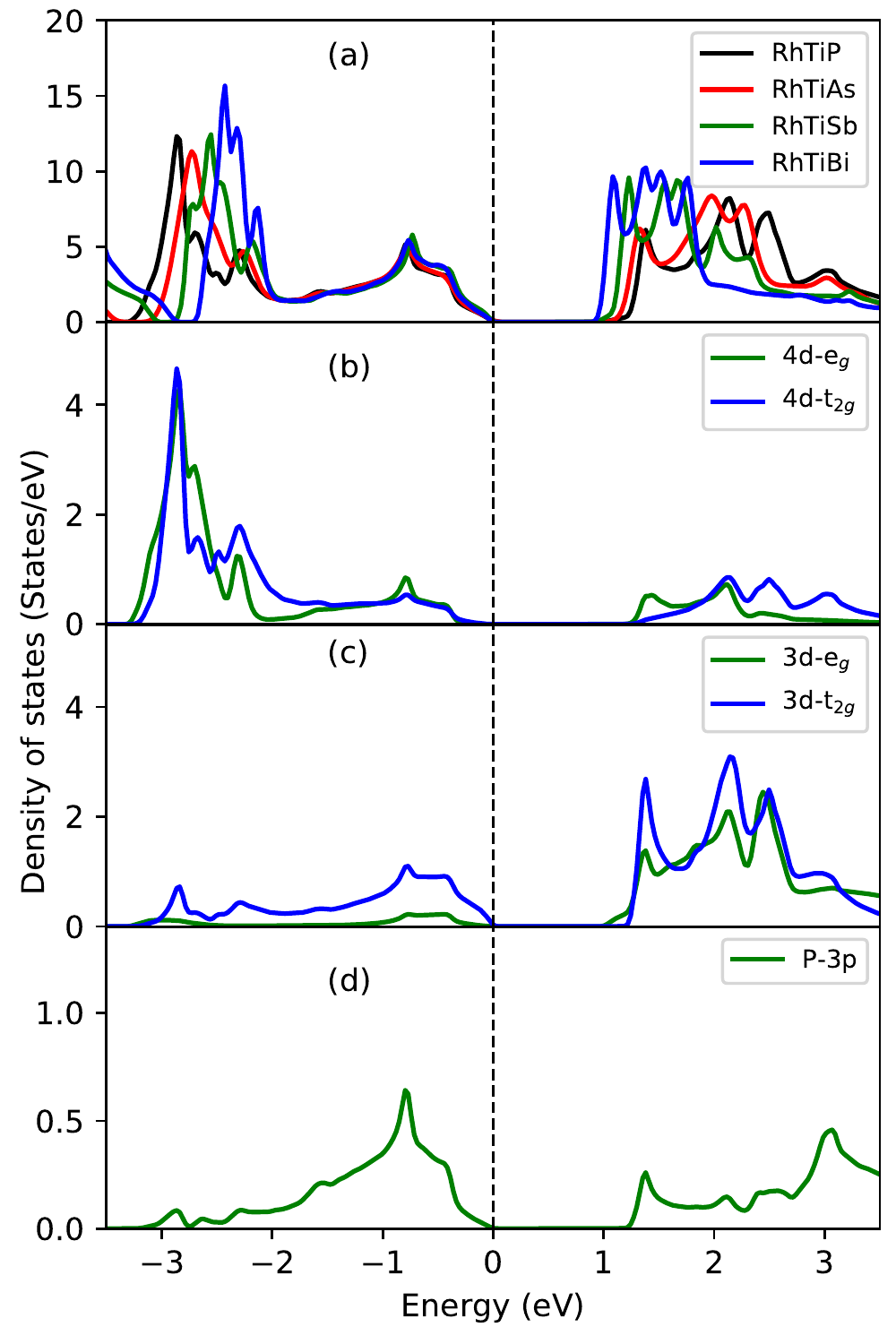,height=4.5in}
	\caption{(a) Total DOS for RhTiZ (Z = P, As, Sb, Bi) and (b)-(d) Partial DOS of Rh-4$d$, Ti-3$d$, and P-3$p$ orbitals in RhTiP within mBJ. Vertical dotted line represent $E_{\rm F}$.} 
	\label{dos}
\end{figure}

Our calculation reveals that these compounds are indirect band gap semiconductors having CBM at $X$- high symmetry point and valence band maximum (VBM) at $\Gamma$. The VBM at $\Gamma$ comprises of light and heavy band with 3-fold degeneracy shown in Figure \ref{band}. For the TE properties, this feature plays significant role. Here, we also calculated the effective mass of the charge carriers and the calculated values of effective mass are shown in the Table \ref{tab3}.
To know more about the nature of Fermi surface in RhTiZ, we computed the iso-energy surfaces around VBM and CBM as shown in Figure \ref{band}(e)-(g). Figure \ref{band}(e) represents the iso-energy surface for RhTiBi at 4 meV below the VBM. Similar features are observed also for RhTiP, RhTiAs, and RhTiSb. Figure \ref{band}(f) represents the iso-energy surface plot for RhTiAs at 4 meV above the CBM, whose features are found similar to RhTiP and RhTiSb. Unlike others, the iso-energy surface of RhTiBi near CBM is noticeably different. This results the band shifts from $E_{\textrm F}$ significantly towards the CBM in RhTiBi. The additional energy surface observed in RhTiBi increasess the electronic density of states (DOS) (see Figure \ref{dos}(a)). This is expected to increase the thermopower i.e. the magnitude of Seebeck coefficient. Thus, we expect to get comparatively better values of Seebeck coefficient for RhTiBi.
 \begin{table}[!h]
	\caption{The effective mass of charge carriers in valence band and conduction band of RhTiZ (Z = P, As, Sb, Bi) in unit of electronic rest mass (m$_e$). The notation ($\Gamma$-K$\Gamma$) represents the effective mass of charge carriers at $\Gamma$ point along $\Gamma$K direction.}
	\label{effM-table}
	\small
	\begin{adjustbox}{angle=0}
			\resizebox{\columnwidth}{!}{
		\begin{tabular}{ l l l l l l l l c l l l c}
			\hline\noalign{\smallskip}
			& & & & &  \multicolumn{7}{c}{Effective mass (m$^*$/m$_e$)}\\
			\cline{6-13} \noalign{\smallskip}
			High-symmetry point & & &    & &RhTiP & & RhTiAs & & RhTiSb & & RhTiBi\\\noalign{\smallskip}
			\hline
			\hline\noalign{\smallskip}
			\textbf{Valence Band ($\Gamma$-K$\Gamma$)} & & &  & & & &  & & & & \\ \noalign{\medskip}
			HH1 & &  &  & & 5.29 & & 5.04 & & 6.62 & & 4.87 \\ \noalign{\smallskip}
			HH2 & &  &  & & 0.66 & & 0.72 & & 0.86 & & 0.85  \\ \noalign{\smallskip}
			LH & & &   & & 0.31 & &  0.30 & & 0.35 & & 0.30 \\ \noalign{\smallskip}
			\hline \noalign{\medskip}
			\textbf{Conduction band (X-$\Gamma$X)} & & & &  & & & &  & & & & \\ \noalign{\medskip}
			Electron & & &  & & 1.03& &  1.21 & & 1.49 & & 1.82 \\ \noalign{\smallskip}
			\hline \noalign{\smallskip}
			\textbf{Conduction band (X-XL)} & & & &  & & & &  & & & & \\ \noalign{\medskip}
			Electron & & &  & & 0.62 & &  0.66 & & 0.68 & & 0.61 \\ \noalign{\smallskip}
			\hline \noalign{\smallskip}
		\end{tabular}}
	\end{adjustbox}
\label{tab3}
\end{table}
Figure \ref{dos} represents the total and partial DOS. The region below the $E_{\textrm F}$ is found to be contributed mainly by the Rh$-4d$ ( $d-t_{2g}$ and $d-e_g$ orbitals), hybridizing with the Ti$-3d$ and Z$-p$ states, while above $E_{\textrm F}$ the major contributions are from the Ti$-3d$ and Z$-p$ states. This suggests that the carrier concentration can be tuned near $E_{\textrm F}$ by doping on to the Z--site.

\subsection{Optical Properties}
\begin{figure}[tbp!]
	\centering
	\psfig{figure=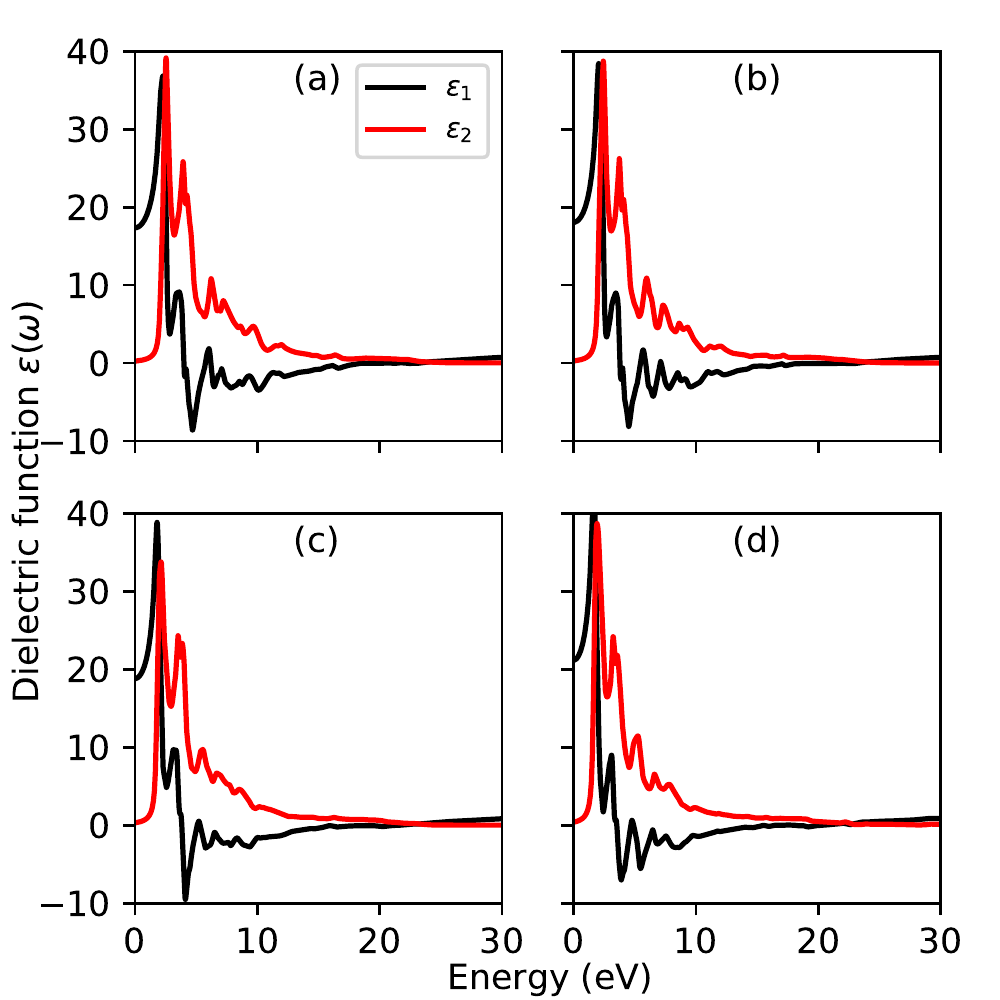,height=3.2in}
	\caption{Calculated real and imaginary parts of dielectric function ($\epsilon_1$ and $\epsilon_2$) as function of incident photon energies for (a) RhTiP, (b) RhTiAs, (c) RhTiSb, and (d) RhTiBi within mBJ.}
	\label{dielectric}
\end{figure}

To explore the optical properties, we calculated the real and imaginary parts of dielectric function, refractive index, reflectivity, absorption spectra and optical conductivity of the hH systems up to the photon energy of 30 eV to point out the response of the materials in solar and high energy radiations. 
The real and imaginary parts of the dielectric function, represented by $\epsilon_1$ and $\epsilon_2$ are shown in Figure \ref{dielectric}. For all the four compounds, there are two major peaks within the photon energy 0 -- 5 eV. The magnitude of the peaks are comparable in all four cases. The obtained static dielectric functions $\epsilon_1(0)$ for RhTiP, RhTiAs, RhTiSb, and RhTiBi are 17.3, 18.0, 18.8, and 21.2, respectively. The imaginary part of the dielectric function shows similar trend for all the alloys. For photon energies higher than 25 eV, the imaginary part of dielectric function tends to zero. This indicates that the radiations  with higher energies transmit through the materials without significant loss of energy. 

\begin{figure}[tbp!]
	\psfig{figure=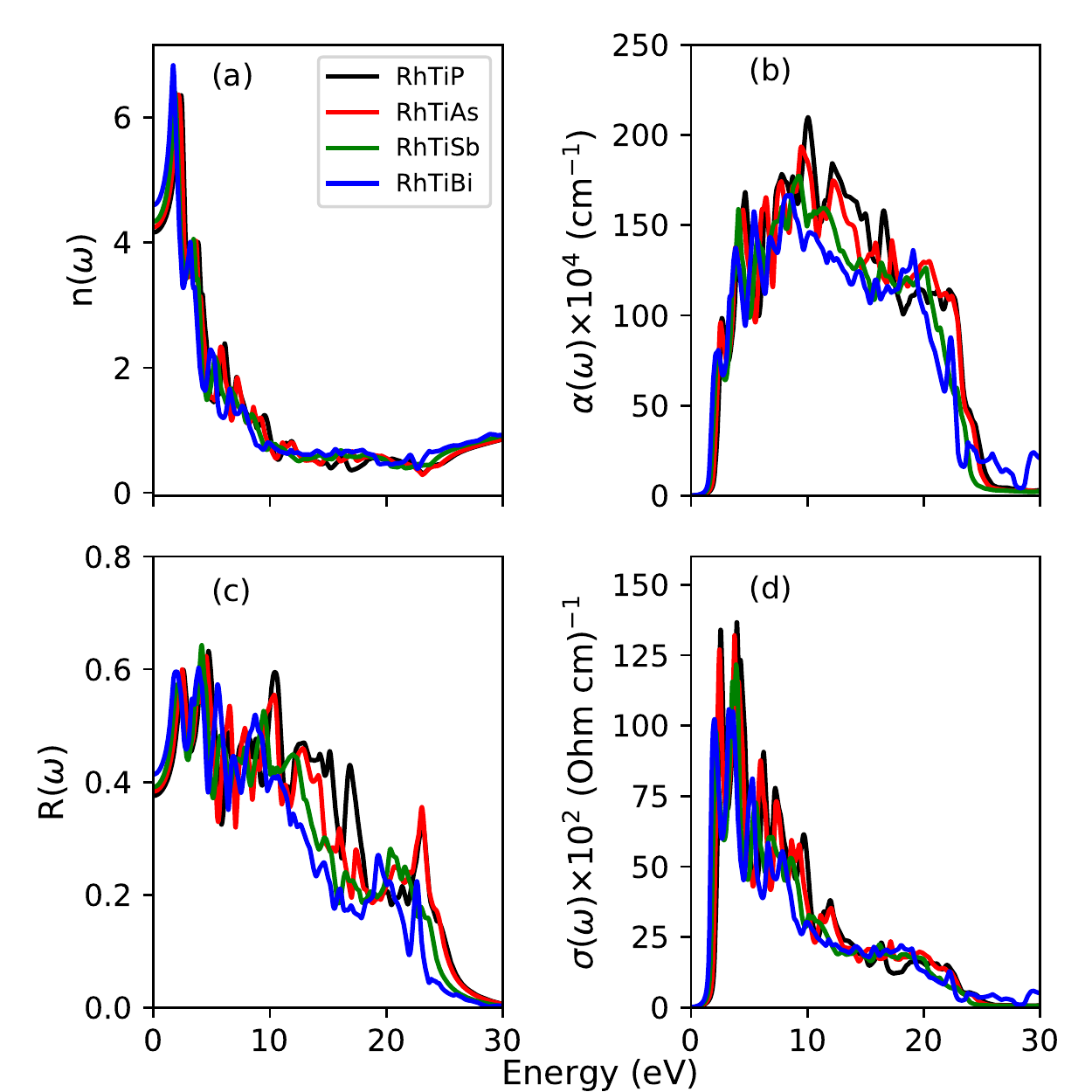,width=\columnwidth} 
	\caption{Calculated (a) refractive index, (b) optical absorption, (c) reflectivity, and (d) optical conductivity as function of incident photon energies for RhTiZ (Z = P, As, Sb, Bi ) within mBJ.}
	\label{refractiveindex}
\end{figure}

Figure \ref{refractiveindex} shows the optical response of RhTiZ. The real part of the refractive index $n(\omega)$ is found to show similar trend as that of dielectric function. 
The static refractive index $n(0)$ for the studied hHs are found to lie within $4.2-4.6$, while the maximum values of the energy dependent refractive index obtained are 6.35 at photon energy 2.4 eV for RhTiP, 6.36 at 2.1 eV for RhTiAs, 6.40 at 1.8 eV for RhTiSb, and 6.83 at 1.7 eV for RhTiBi, respectively. 

It is well-known that the optical absorption coefficient signifies the thickness of the material up to which the incident light can penetrate before being completely absorbed. It gives information on the solar energy conversion efficiency which is important for application as optoelectronic devices. The calculated absorption spectra is shown in Figure \ref{refractiveindex} (b). The absorption peaks are found around 2 eV. Despite the replacement of P by other elements, the effect on the magnitude and position of the absorption spectra are negligible. 

In the visible region, RhTiP has the smallest reflectivity out of the four compounds under study, and is found to increase as P is replaced by As, Sb, and Bi (see Figure \ref{refractiveindex} (c)). These alloys have the maximum reflectivity for photon energies within 3--5 eV. In addtion, the real part of the optical conductivity have been calculated (see Figure \ref{refractiveindex} (d)). At low energy region (i.e., $<$ 8 eV), high optical conductivity has been noticed. The values are comparable with other hH alloys.\cite{anissa2019optical,dey2020extensive}

 \begin{figure}[!t]
	\centering
	\psfig{figure=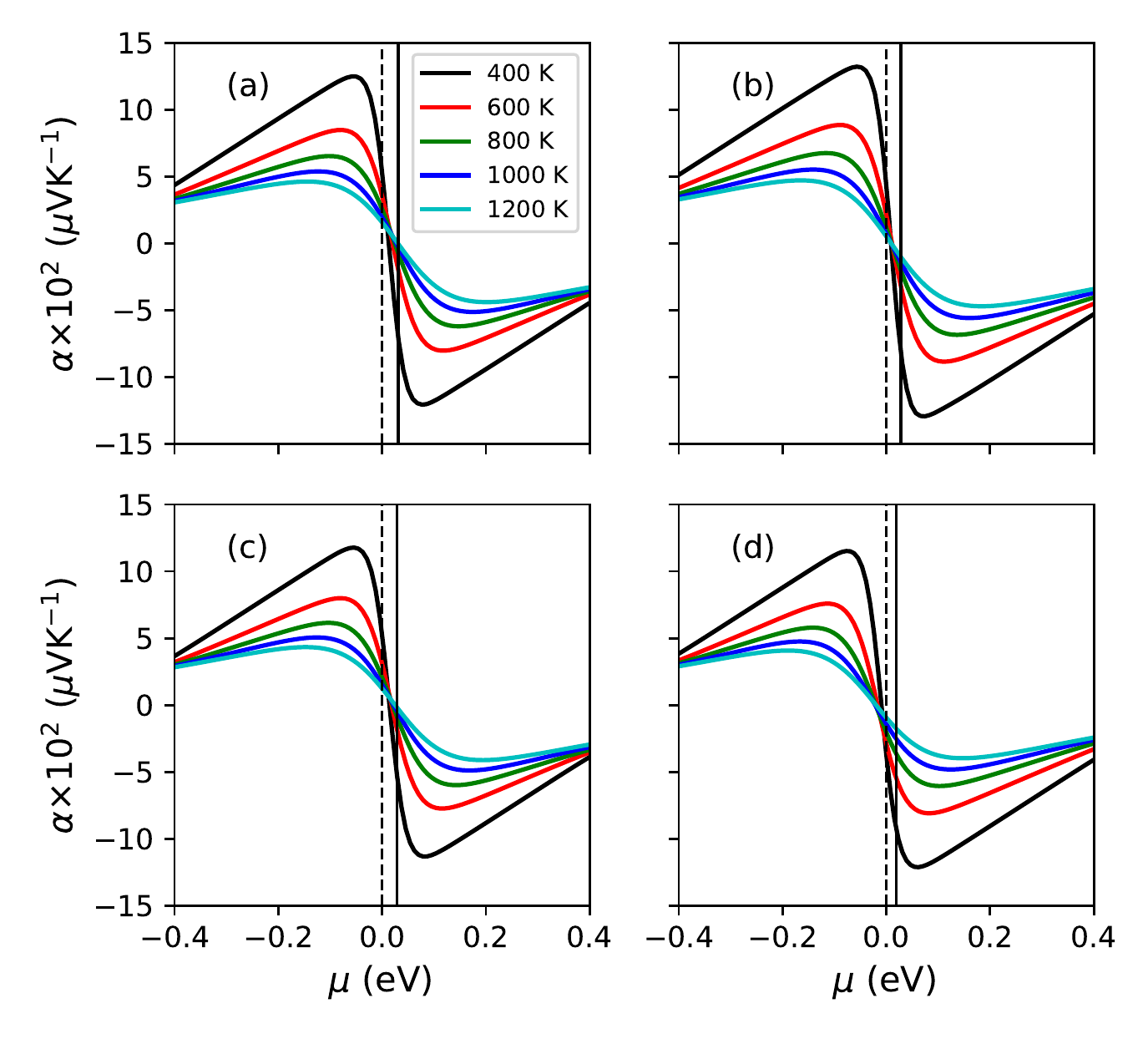,width=3in, height=2.8in}
	\caption{Variations of Seebeck coefficient with chemical potential for (a) RhTiP, (b) RhTiAs, (c) RhTiSb, and (d) RhTiBi at different temperatures. The vertical dotted lines are the respective middle of band gap, and the solid vertical lines represent the chemical potential at 300 K.}
	\label{S-v-mu}
\end{figure}

\subsection{Transport Properties}
The suitability of a material for TE purpose depends on properties like Seebeck coefficient ($\alpha$), electrical conductivity ($\sigma$) and thermal conductivity ($\kappa$). A factor that improves one of these properties can adversely affect another. For instance, a heavy conduction or valence band is good for Seebeck coefficient but usually results in smaller electrical conductivity. Therefore, it is necessary to optimize all these properties to maximize the figure of merit $ZT$. In this work, we have studied $\alpha$, $\sigma$, and $\kappa$ of considered hH compounds using constant relaxation time approximation (CRTA) and rigid band approximation (RBA).
 
The magnitude of $\alpha$ is dependent upon the density of state effective mass, which is directly proportional to the energy valley degeneracy in the band structure and the effective mass of the charge carriers. To estimate the values of $\alpha$ for the RhTiZ compounds, we  calculated the effective mass of charge carriers using the electronic band structure. The effective mass is decided by the curvature of the energy band. The flat bands in the band structure corresponds to larger effective mass. 
In case of RhTiZ, VBM is threefold degenerate and lies at the $\Gamma$. The CBM lies at the high symmetric point $X$. An interesting observation is that, as we move down the group from P to Bi, the band above the CBM shifts down, and in case of RhTiBi, the conduction band is also twofold degenerate at $X$ point. We have estimated the effective mass of holes at the $\Gamma$ along $K\Gamma$, and the effective mass of electrons at $X$ along $XL$ and $\Gamma X$. The calculated effective masses are listed in Table \ref{effM-table}. Here, the notation $\Gamma-K\Gamma$ implies that the effective mass was calculated at $\Gamma$ along the $K\Gamma$. The valence bands are labeled as HH1, HH2 and LH as shown in Figure \ref{band}(a). 
The effective mass of holes in the heaviest valence band HH1 is much larger than the effective mass of electrons in the conduction band. The effective mass for LH bands are comparable among four hH alloys. The maximum value of effective mass for holes in HH1 at $\Gamma$ along K$\Gamma$ is 6.62 m$_e$ for RhTiSb followed by RhTiP (5.29 m$_e$ ), RhTiAs (5.04 m$_e$), and RhTiBi (4.87 m$_e$), respectively. Furthermore, the maximum value of effective mass for electron at $X$ along $\Gamma X$ is 1.82 m$_e$ for RhTiBi, whereas the lowest value achieved at $X$ along $\Gamma X$ is 1.03 m$_e$ for RhTiP.

From the above calculation on the effective mass, the maximum values noted are for holes. This suggests that holes will contribute largely to thermopower. Thus, these materials are expected to have positive value of $\alpha$. It should however be noted that the chemical potential $\mu$ for an intrinsic semiconductor depends on the temperature and the effective masses of holes and electrons following the relation:\cite{KittelBook}
 \begin{equation}
	\mu = \epsilon_\nu + \frac{1}{2} E_g + \frac{3}{4}\kappa_B T ln(m_v/m_c)
\end{equation}
 where $\epsilon_\nu$ is the energy of the electron at the VBM, $m_v$ and $m_c$ are the effective mass of the charge carriers in the valence band and conduction band respectively, and $E_g$ is the energy band gap.

\begin{figure}[t]
	\psfig{figure=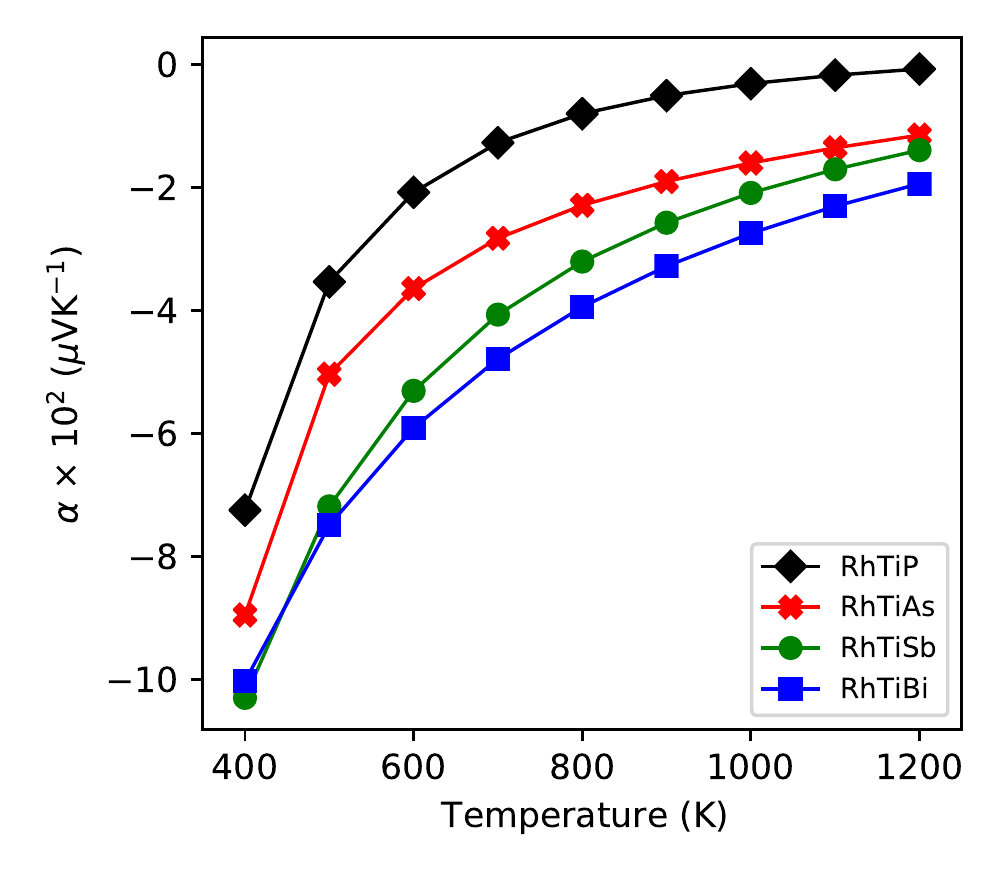,width=3in, height=2.8in}
	\caption{Variations of Seebeck coefficient with temperature for RhTiP, RhTiAs, RhTiSb, and RhTiBi.}
	\label{S-v-T}
\end{figure}
\begin{figure}[h]
	\psfig{figure=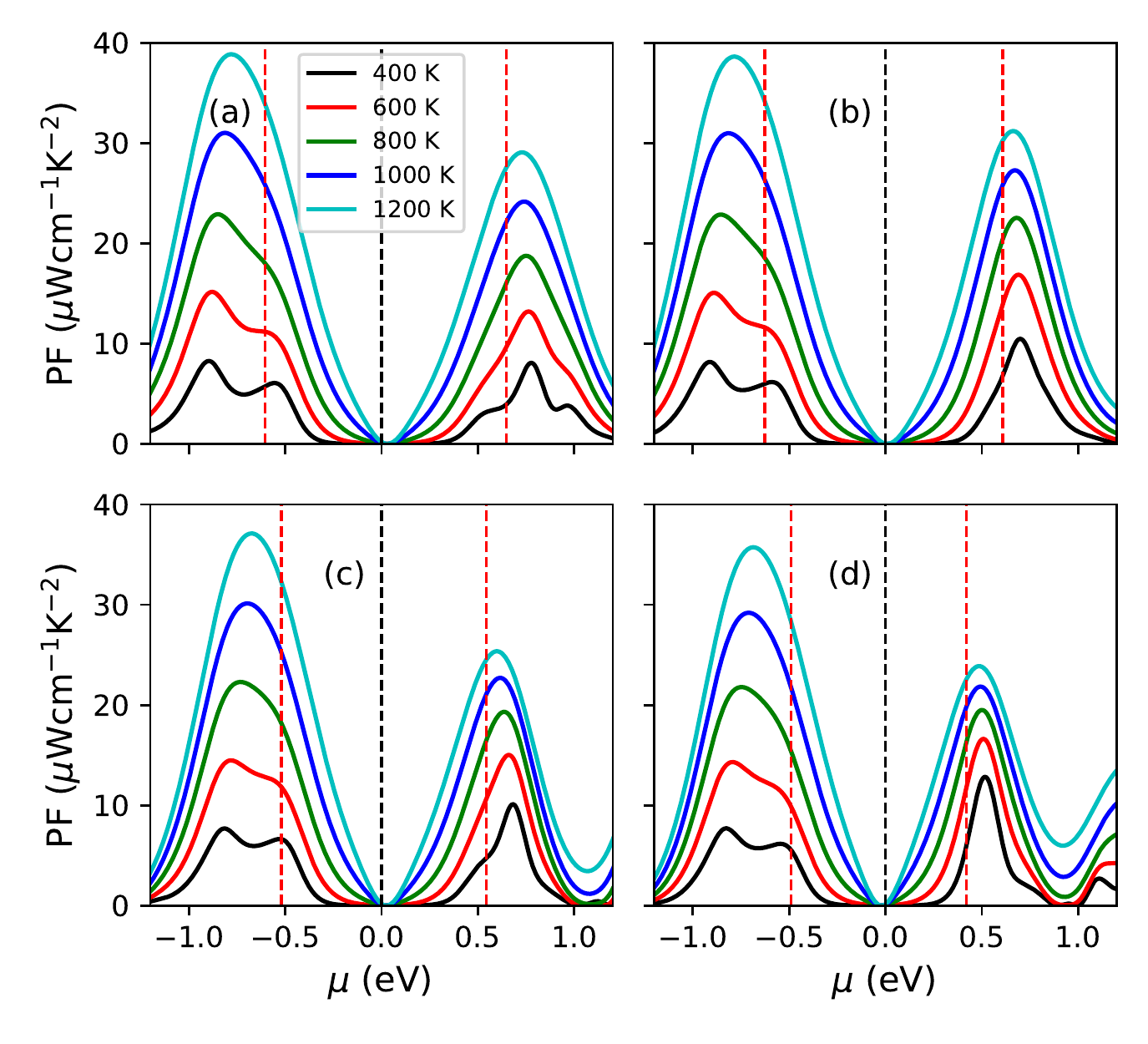,width=3in, height=2.8in}
	\caption{Variations of power factor with chemical potential for (a) RhTiP, (b) RhTiAs, (c) RhTiSb, and (d) RhTiBi at different temperatures. The vertical dotted lines at positive (negative) $\mu$ corresponds to the optimal value for $ZT$ of $n$--type ($p$--type) doping cases, and the vertical line at $\mu$=0 represent middle of the band gap.}
	\label{PF-v-mu}
\end{figure}
Using the effective mass for HH1 band and the effective mass of electron along $\Gamma X$ direction from Table \ref{effM-table} for $m_v$ and $m_c$ respectively, we calculated the value of chemical potential $\mu$ at 300 K. In all the cases, the chemical potential is shifted towards the conduction band. The chemical potential at 300 K was 31 meV, 28 meV, 29 meV, and 19 meV above the middle of the band gap for RhTiP, RhTiAs, RhTiSb, and RhTiBi, respectively. The variation of $\alpha$ with $\mu$ at different temperatures are shown in Figure \ref{S-v-mu}. 
The dashed (solid) vertical lines represents the middle of band gap (chemical potential at 300 K). The calculated values of $\alpha$ in the temperature range 400 -- 1200 K for a fixed $\mu$ for 300 K are shown in Figure \ref{S-v-T}. All the hH alloys have negative Seebeck coefficients that decrease in magnitude as temperature is increased. The largest values of $\alpha$ are observed for RhTiBi, which can  be attributed to the two-fold degeneracy of the conduction band at the CBM in the bandstructure of RhTiBi. At 400 K, the values of $\alpha$ are -727 \seebeckunit, -896 \seebeckunit, -1030 \seebeckunit, and -1000 \seebeckunit for RhTiP. RhTiAs, RhTiSb, and RhTiBi, respectively. The decrease in the magnitude of $\alpha$ with increase in temperature could be due to the rise in the carrier concentration.
 
To determine the optimum doping level for enhancing the TE performance, the power factor (PF) has been calculated for all four systems. Due to the unavailability of experimental value of relaxation time of charge carriers, $\tau$, we have used the CRTA and fixed the value of $\tau$ at 2$\times$10$^{-15}$ s. This value was chosen based on our previous work,\cite{dhurba2021} where we estimated $\tau$ for NiTiSn by using the experimental values of electrical conductivity from Kim $et$ $al$.\cite{kim2007high} The variations in the PF with respect to $\mu$ for different temperatures are shown in Figure \ref{PF-v-mu}. Maxima are noted in the values of PF for both positive and negative values of $\mu$. 
Positive region of $\mu$ represents the $n$--type doping, while the negative represents $p$--type doping. Up to the temperature of 600 K, the maximum values of PF for $n$--type and $p$--type are comparable. The maximum values of PF obtained at 600 K for $p$ ($n$)--type are  15.16(13.20), 15.08 (16.87), 14.48 (15.03), and 14.27 (16.65) \PFunit\ for RhTiP, RhTiAs, RhTiSb, and RhTiBi, respectively. 

These values are comparable to the reported PF for some cobalt--based hH materials. \cite{zeeshan2017Co} For temperatures larger than 600 K, the maximum PF values achievable by $p$--type doping are significantly larger than those obtainable by $n$--type doping. Hence, for high temperature region (T > 600 K), $p$--type doping is more preferable for enhancing the value of PF.
\begin{figure}[!h]
	\psfig{figure=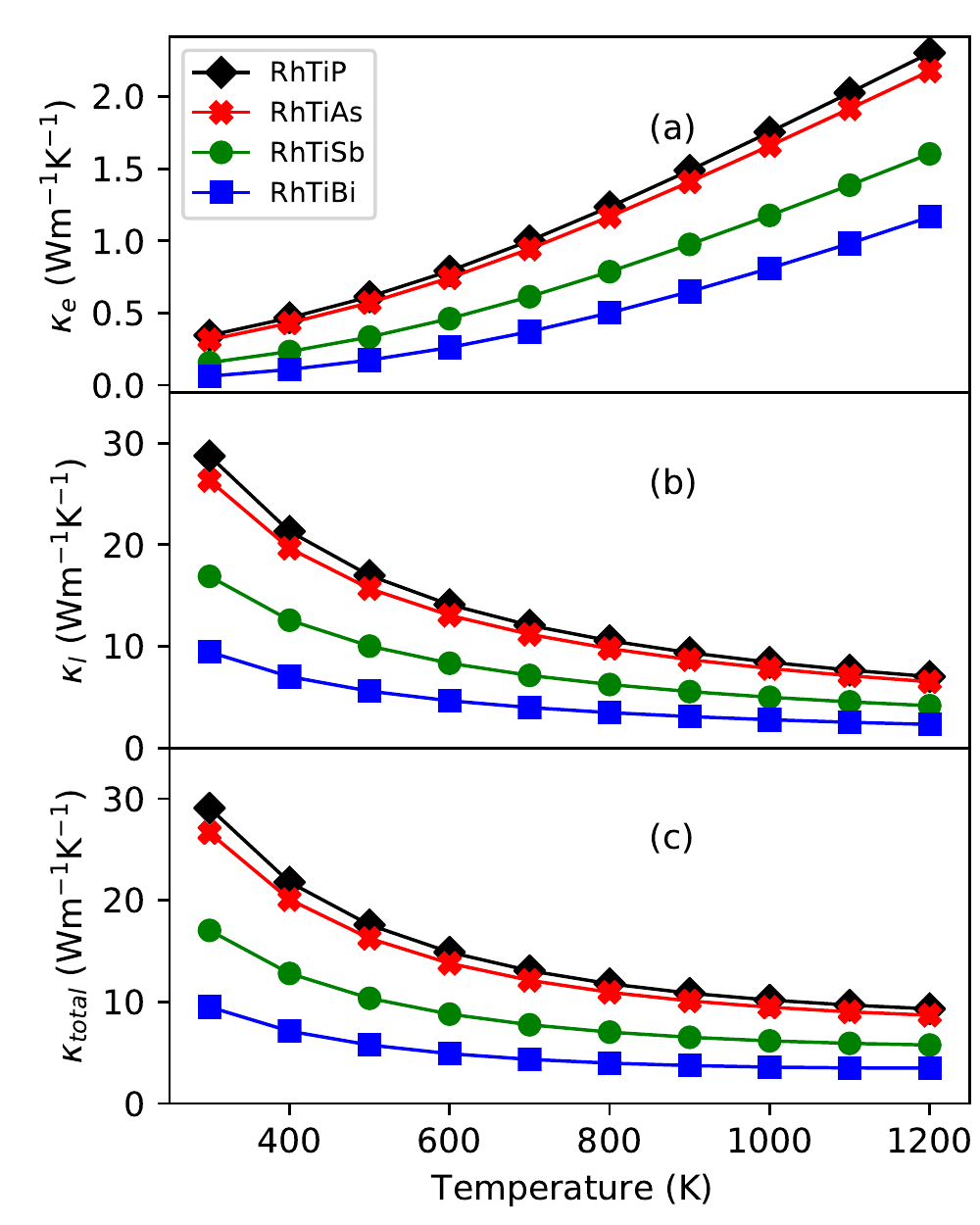,height=3.5in}
	\caption{(a) Calculated electronic thermal conductivity ($\kappa_e$), (b) Calculated lattice thermal conductivity ($\kappa_l$), and the total thermal conductivity ($\kappa_{total}$) as a function of temperature for  RhTiP, RhTiAs, RhTiSb, and RhTiBi.}
	\label{thermalconduc}
\end{figure}
\begin{figure*}[bth!]
	\psfig{figure=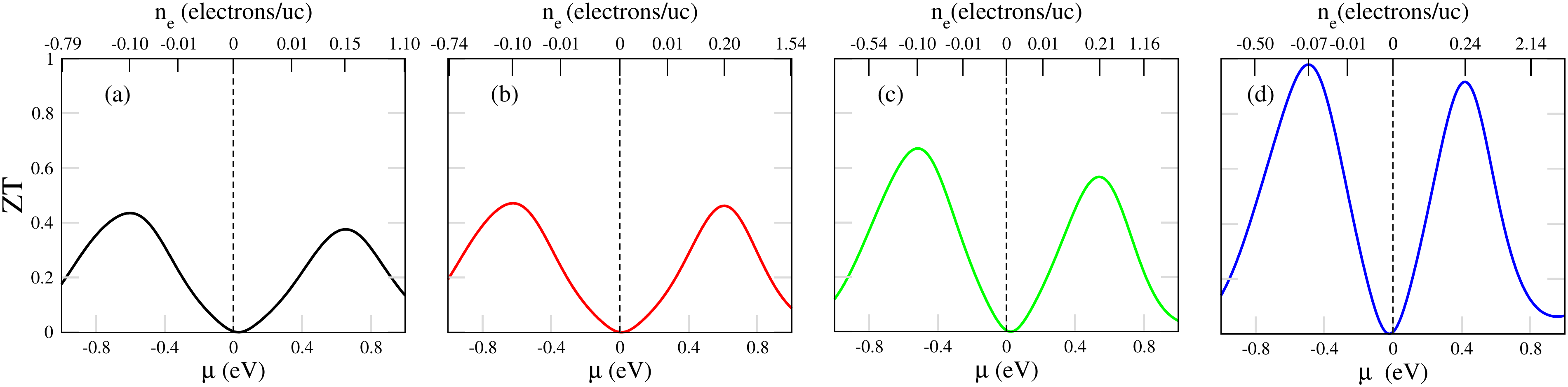,width=\textwidth}
	\caption{Variations of $ZT$  with chemical potential for (a) RhTiP, (b) RhTiAs, (c) RhTiSb, and (d) RhTiBi at 1200 K. The upper horizontal axis denotes the number of electrons per unit cell (uc), and the negative value corresponds to hole doping.}
	\label{ZT-v-mu-n}
\end{figure*}
\begin{table}[h!]
	\caption{Calculated $n$--type and $p$--type doping levels and the corresponding Seebeck coefficient, electrical conductivity, PF, thermal conductivity, and	$ZT$ of RhTiZ ( Z= P, As, Sb, Bi) in cubic symmetry $F\bar{4}3m$ at 1200 K. Here, negative values of n$_e$ correspond to $p$--type doping, and the positive n$_e$ correspond to $n$--type doping.}
	\begin{tabular}{llllll}
		\hline \noalign{\smallskip}
		&          & RhTiP & RhTiAs & RhTiSb & RhTiBi \\ \hline \hline \noalign{\smallskip}
		\multirow{6}{*}{$n$--type} & n (e/uc) &0.15&0.20&0.21&0.24 \\ \noalign{\smallskip}
		&  $\alpha$ (\seebeckunit) &-183.25&-200.40&-202.20&-228.61 \\ \noalign{\smallskip}
		&  $\sigma$$\times$10$^4$ (\sigmaunit)	&8.22&7.53&6.00&4.32  \\ \noalign{\smallskip}
		&  $\alpha^2 \sigma$ (\PFunit)	&27.60&30.24&24.51&22.60 \\ \noalign{\smallskip}
		&  $\kappa$ (\kappaunit) 		&8.81&7.85&5.19& 2.96 \\ \noalign{\smallskip}
		&  $ZT$       					&0.38&0.46&0.58&0.91 \\ \hline \noalign{\smallskip}
		\multirow{6}{*}{$p$--type} & n (e/uc)  &-0.10&-0.10&-0.10&-0.074  \\ \noalign{\smallskip}
		&  $\alpha$ (\seebeckunit)		&188.08&191.54&210.58&232.71  \\ \noalign{\smallskip}
		&  $\sigma$ (\sigmaunit)  		& 9.57&9.31&7.27&5.24  \\ \noalign{\smallskip}
		&  $\alpha^2 \sigma$ (\PFunit) 	&33.85&34.15&32.22&28.39  \\ \noalign{\smallskip}
		&  $\kappa$ (\kappaunit)		&9.32&8.69&5.76&3.48  \\ \noalign{\smallskip}
		&  $ZT$       					&0.44&0.47&0.67&0.98  \\ \hline
	\end{tabular}
	\label{tab4}
\end{table}
\begin{figure*}[ht!]
	\psfig{figure=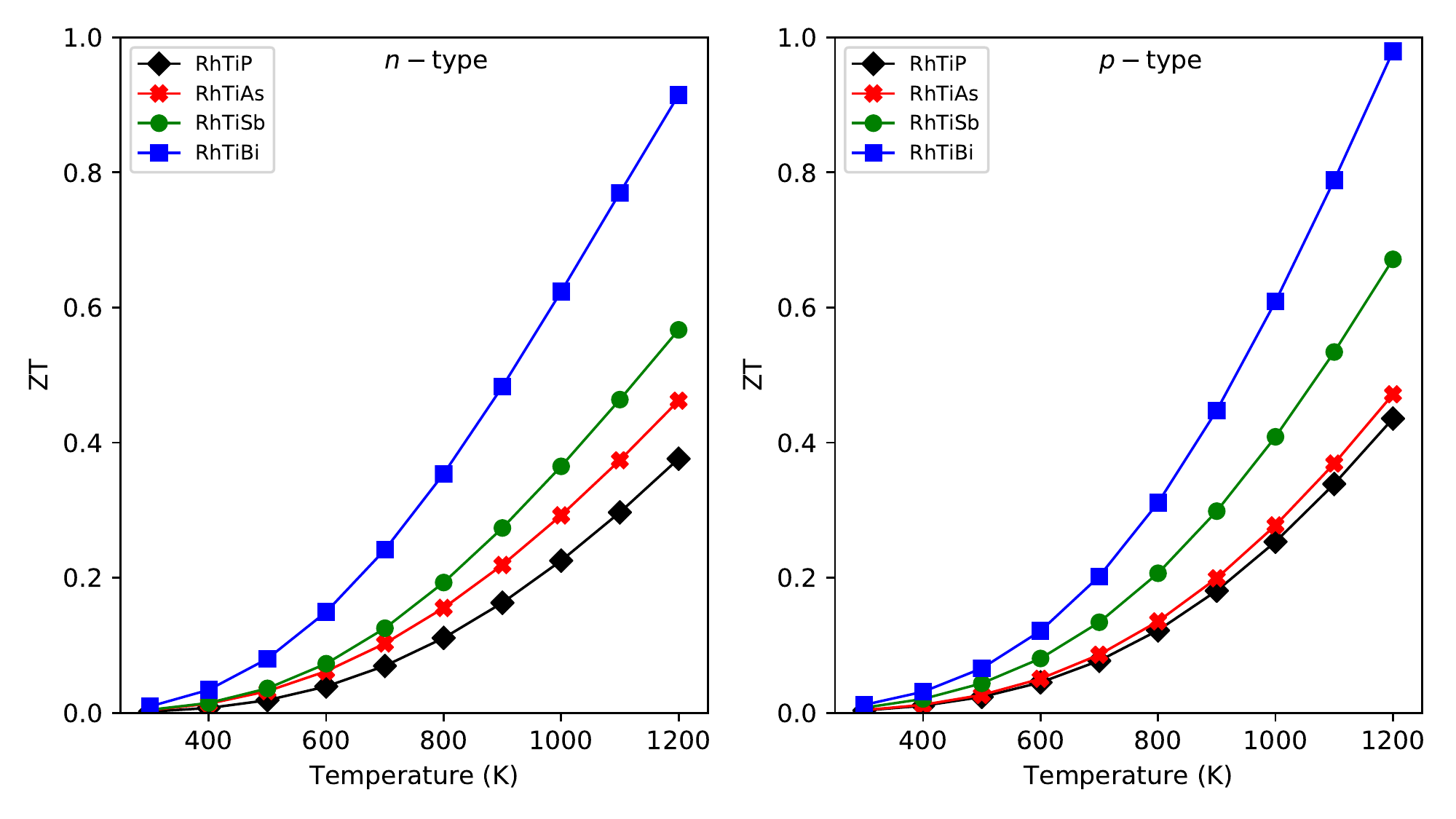,width=5in, height=2.5in}
	\caption{The temperature-dependent variation of $ZT$ values for $p$--type and $n$--type RhTiP, RhTiAs, RhTiSb, and RhTiBi.} 
	\label{zt}
\end{figure*}
The optimum $p$--type ($n$--type) doping levels for RhTiP, RhTiAs, RhTiSb, and RhTiBi are 0.30 (0.28), 0.28 (0.31), 0.29 (0.34), and 0.27 (0.40) holes (electrons) per unit cell, respectively. 

In all cases, the optimum PF values for $p$--type doping are significantly larger than that for $n$--type doping. The additional advantage of $p$--type doping is that, the better value of PF is achievable by smaller doping concentrations. However, the doping levels with optimum PF do not necessarily correspond to optimum \zt \ values.
  
The thermal conductivity of the materials at different temperatures are shown in Figure \ref{thermalconduc}. 

The calculated values of the electronic thermal conductivity $\kappa_e$ (lattice thermal conductivity $\kappa_l$) are 2.30 (7.02), 2.17 (6.51), 1.60 (4.16), and 1.15 (2.31) Wm$^{-1}$K$^{-1}$ for RhTiP, RhTiAs, RhTiSb, and RhTiBi respectively at 1200 K (Figure \ref{thermalconduc}(a and b)). 
The values of $\kappa_e$ reported here corresponds to the optimum $ZT$ with $p$--type doping. As can be seen in Figure \ref{thermalconduc}(b), $\kappa_l$ is found to decrease with increase in temperature for all cases. Note that the values reported here for $\kappa_l$ are calculated for undoped RhTiZ. These values were used for the calculation of $ZT$ for different doping levels. Among the studied alloys, the $\kappa_l$ value for RhTiP is highest, and is found to decrease as P is replaced by As, Sb, and Bi. 
This might be due to the scattering of phonons from the heavier atoms.\cite{fu2015realizing, fu2016} 
Due to very small value of $\kappa_l$, RhTiBi has the largest $ZT$ value among the studied alloys. 

Figure \ref{ZT-v-mu-n} shows the variation of \zt \ with the chemical potential and doping levels at 1200 K. The doping level corresponding to optimum PF (see Figure\ref{PF-v-mu}) are different when compared to those corresponding to optimum \zt \  values (see Figure \ref{ZT-v-mu-n}). 
The different values of TE parameters corresponding to $n$ and $p$--type doping at 1200 K are summarized in the Table \ref{tab4}. 
The variation of $ZT$ with temperature for $n$ and $p$--type doping are shown in Figure \ref{zt}. The optimum $ZT$ values of  $p$--type ($n$--type) RhTiP, RhTiAs, RhTiSb, and RhTiBi are 0.44 (0.38), 0.47 (0.46), 0.67 (0.58), and 0.98 (0.91), respectively at 1200 K. These values of $ZT$ are lower than those for the state-of-the-art TE materials,\cite{snyder2008, fu2015realizing, zhu2019discovery} but are comparable to several other works done on Heusler alloys.\cite{kim2007high,dhurba2021,hu2020electronic}
Thus, these Rh-based hH alloys are suitable candidates for use as both $p$--type and $n$--type elements in TE devices.
 		
\section{CONCLUSIONS}
We performed the density functional theory calculations to explore the structural, electronic, mechanical, optical, and thermoelectric properties of rhodium based hH compounds RhTiP, RhTiAs, RhTiSb and RhTiBi. All these alloys are found to be semiconducting in nature. In case of RhTiBi, an extra electron pocket is observed at the CBM. The Pugh's ratio calculation indicates that RhTiP is ductile while the remaining alloys are brittle in nature. Optical study reveals that RhTiZ (where Z are P, As, Sb, and Bi) have high absorption coefficient and optical conductivity, due to which these alloys are predicted to be suitable for optoelectronic applications. We calculated the power factor which suggests that $p$--type doping is more favorable to enhance the TE properties. The obtained optimum $ZT$ values are 0.38, 0.46, 0.58 and 0.91 for $n$--type and 0.44, 0.47, 0.67, and 0.98 for $p$--type RhTiP, RhTiAs, RhTiSb and RhTiBi, respectively at 1200 K. This indicates that the alloys are good candidates for use in TE devices. Brittleness can be an issue during synthesis and use in TE devices in few systems. To address this, doping and annealing may be possible ways to improve ductility. Furthermore, the value of $\tau$ used in this study might be larger than the actual value of the relaxation time. In that case, the \zt\ values might be smaller than the values calculated here. Hence, an experimental study of the transport properties of these materials can be a very interesting prospect for experimentalists.

\section{acknowledgments}
D. R. Jaishi thanks Rajendra Adhikari for fruitful discussion and suggestions. Most of the computational works was performed at the Advanced Materials Research Laboratory (AMRL), Central Department of Physics, Kirtipur, Tribhuvan University supported by Alexander von Humboldt Foundation, Germany as equipment grants. 
\bibliography{Reference.bib}
\end{document}